# MgB$_2$ superconducting thin films with a transition temperature of 39 Kelvin

W. N. Kang[*], Hyeong-Jin Kim, Eun-Mi Choi, C. U. Jung, and Sung-Ik Lee

*National Creative Research Initiative Center for Superconductivity, Department of Physics, Pohang University of Science and Technology, Pohang 790-784, Korea*

We report the growth of high-quality c-axis-oriented epitaxial MgB$_2$ thin films by using a pulsed laser deposition technique. The thin films grown on ($1\ \bar{1}\ 0\ 2$) Al$_2$O$_3$ substrates show a T$_c$ of 39 K. The critical current density in zero field is ~ 6 x 10$^6$ A/cm$^2$ at 5 K and ~ 3 x 10$^5$ A/cm$^2$ at 35 K, suggesting that this compound has great potential for electronic device applications, such as microwave devices and superconducting quantum interference devices (SQUIDs). For the films deposited on Al$_2$O$_3$, X-ray diffraction patterns indicate a highly c-axis-oriented crystal structure perpendicular to the substrate surface.

The recent discovery of the binary metallic MgB$_2$ superconductor [1] having a remarkably high transition temperature (T$_c$) of 39 K has attracted great scientific interest [2-8]. With its metallic charge carrier density [2] and the strongly linked nature of the inter-grains in a polycrystalline form [9, 10], this material is expected to be a very promising candidate for superconducting device [11] as well as large-scale applications. Furthermore, since the single crystal growth of MgB$_2$ seems very difficult, the fabrication of epitaxial thin film should be an important development for future basic research studies. However, the fabrication of thin films of this material has not been reported yet.

We used a two-step method to fabricate MgB$_2$ thin films. First, we deposited amorphous B thin films; we sintered then at high temperature in Mg vapor, which is very similar to the growth techniques of cuprate Hg-based superconducting thin films [12, 13]. We pressed commercial Boron (99.99%) powder into a disk shape with a diameter of 12.7 mm and a height of 5 mm under a pressure of 6 tons. Precursor thin films of B were deposited on Al$_2$O$_3$ (AO) and SrTiO$_3$ (STO) substrates at room temperature by using pulsed laser deposition. The laser energy density was 20 – 30 J/cm$^2$ at a laser flux of 600 mJ/pulse and a pulse frequency of 8 Hz. After a precursor thin film had been fabricated, it was put into a Ta tube together with a high purity Mg metal (99.9%) and sealed in an Ar atmosphere. The heat treatment was carried out in an evacuated quartz ampoule to prevent oxidation of the Ta tube. The typical sintering procedure was fast heating to 900 °C in 5 minutes; this temperature was held for 10 - 30 minutes, and then quenched to room temperature. The typical film thickness used in this study was 0.4 μm, which was measured by a scanning electron microscope. This simple technique can be applied to other physical deposition methods, such as sputtering and electron-beam evaporation, and is highly reproducible, so mass production should be possible. The resistivity measurements were carried out using the dc four-probe method. The dc magnetic properties were measured with a Quantum Design MPMS superconducting quantum interference device magnetometer. The structures were analyzed using a x-ray diffractometer (XRD).

The typical temperature dependence of the resistivity of MgB$_2$ grown on AO is shown in Fig. 1. The inset is a magnified view near the T$_c$ region. The resistivity of the film begins to enter the superconducting transition at 39 K and goes to zero resistance at 37.6 K. A very sharp transition, with a width of ~ 0.7 K from 90% to 10% of the normal state resistivity, is evident in the inset. This is comparable to most reported values for high-quality bulk samples [3-6]. The normal-state resistivity at 290 K was ~ 4.7 μΩ cm, indicating an intermetallic nature with a relatively high charge carrier density [2]. This resistivity is smaller than those for polycrystalline MgB$_2$ wire [10] and for bulk samples synthesized under high pressure [3, 4]. Most of our films fabricated under the same conditions showed a similar superconducting transition around 39 K.

Fig. 2 shows the zero-field-cooled (ZFC) and the field-cooled (FC) dc magnetization (M) curves of a MgB$_2$ thin film in a 10 Oe field applied parallel to the c-axis. The irreversibility temperature detected at 37.5 K coincides with the zero-resistance temperature obtained from resistivity measurement. The ZFC curve shows a rather broad diamagnetic transition compared to the resistivity data. To estimate the critical current density (J$_c$), we measured the M-H loop for the same sample as a function of temperature, as shown in the inset of Fig. 2. At zero field, the J$_c$ calculated using the Bean model was ~ 6 x 10$^6$ A/cm$^2$ at 5 K and ~ 3 x 10$^5$ A/cm$^2$ at 35 K, which are 10 times higher than the values obtained for a MgB$_2$ wire [10] by using transport measurements and slightly smaller than the values for Hg-based superconducting thin films [12]. It should be noted that we used the sample size rather than the grain size in calculating J$_c$. In view of their strongly linked nature of the inter-grains [9] and their excellent electrical and magnetic characteristics compare to conventional superconductors, these films would be useful in applications of electronic devices, such as microwave devices and portable SQUIDs sensors by using miniature cryocoolers with low power consumption.

Structural analysis was carried out by XRD, which is shown in Fig. 3. The a- and the c-axis lattice constants determined from the (101) and the (00$l$) peaks are observed to be 0.310 and 0.352, respectively. Interestingly, the XRD patterns indicate that the films deposited on (a) AO are epitaxially aligned with the c axis while the films on (b) STO is well aligned with the (101) direction normal to the substrate planes. These results suggest that we may be able to control the orientation of MgB$_2$ thin film simply by using different substrates if the optimum growth condition is explored. We also find that the AO substrates are chemically very stable during the heat treatment at high temperatures in Mg vapor. With its high thermal conductivity and small dielectric constant, Al$_2$O$_3$ is a very promising substrate in terms of superconducting device applications, such as microwave devices and portable sensors. For the STO substrates, however, the chemical reaction between the substrate and Mg was observed and appeared as minor impurity phases, as shown in Fig. 3 (a). Considering that STO has a cubic structure with a lattice constant of 0.389 nm and R-plane AO has $a_0 = 0.476$ nm and $b_0^* = (c_0^2 + 3a_0^2)^{1/2} = 1.538$ nm, it is quite striking that MgB$_2$ thin films grow with preferred orientations on AO and STO substrates even though the lattice-matching relationship between the MgB$_2$ and the substrates is not well satisfied. Further microscopic investigation of the MgB$_2$ crystallographic relations on AO and STO substrates is required.

∗To whom correspondence should be addressed (e-mail: wnkang@postech.ac.kr).

**Figure Captions**

Fig. 1. Resistivity vs. temperature for a MgB$_2$ thin film grown on an Al$_2$O$_3$ substrate by using pulsed laser deposition with post annealing techniques. The inset is a magnified view of the temperature region of 36 - 40 K for the sake of clarity. A sharp superconducting transition is observed at 39 K.

Fig. 2. Magnetization vs. temperature at H = 10 Oe for a MgB$_2$ thin film. The inset shows the M-H hysteresis loop at 5 K (solid circle) and 35 K (open circle). Note that very high current-carrying capabilities of ~ 6 x 10$^6$ A/cm$^2$ at 5 K and ~ 3 x 10$^5$ A/cm$^2$ at 35 K were observed at zero field.

Fig. 3. XRD patterns for MgB$_2$ thin films grown on (a) (100) SrTiO$_3$ and R-plane (b) (1 $\bar{1}$ 0 2) Al$_2$O$_3$ substrates. The (00$l$) peaks of MgB$_2$ grown on Al$_2$O$_3$ indicate c-axis-oriented epitaxial thin films whereas MgB$_2$ deposited on SrTiO$_3$ show (101) plane-oriented thin films.

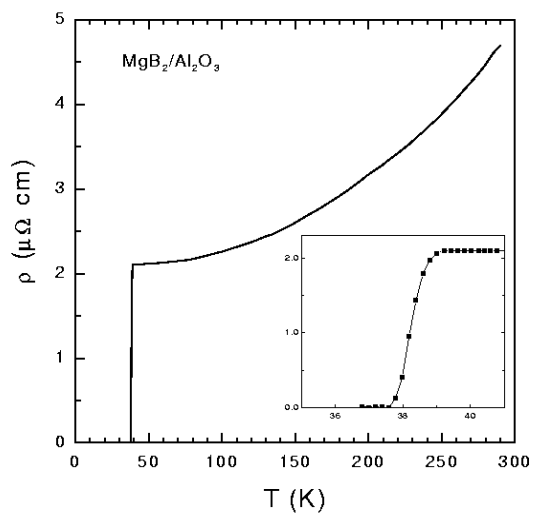

Fig. 1. Kang et al.

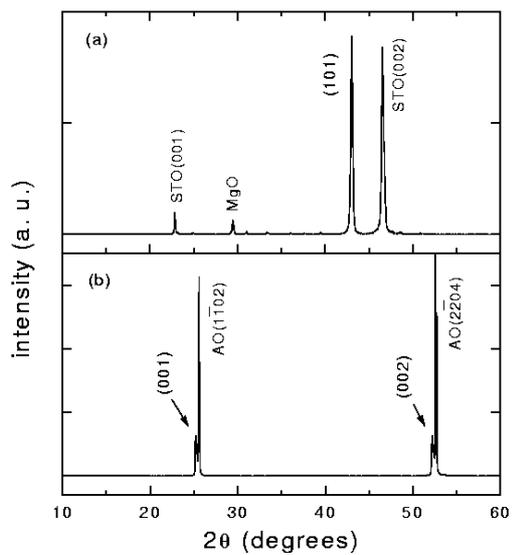

Fig. 3. Kang et al.

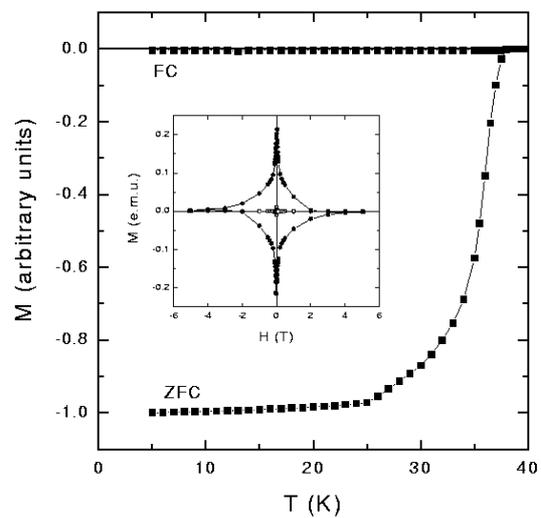

Fig. 2. Kang et al.